\documentclass[nofootinbib,twocolumn,aps,letterpaper,superscriptaddress,showpacs]{revtex4}
\usepackage{amsmath}
\usepackage{amsfonts}
\usepackage{amssymb}
\usepackage[dvips]{graphicx}

\begin{document}

\newcommand\be{\begin{equation}}
\newcommand\ee{\end{equation}}
\newcommand\ba{\begin{eqnarray}}
\newcommand\ea{\end{eqnarray}}
\newcommand\bseq{\begin{subequations}} 
\newcommand\eseq{\end{subequations}}
\newcommand\bcas{\begin{cases}}
\newcommand\ecas{\end{cases}}
\newcommand{\p}{\partial}
\newcommand{\f}{\frac}
\newcommand{\nn}{\nonumber \\}
\def\tr{{\rm Tr}}

\title{Big Bounce in Dipole Cosmology}

\author{Marco Valerio Battisti}
\email{battisti@icra.it}
\affiliation{Centre de Physique Th\'eorique, Case 907 Luminy, 13288 Marseille, EU}
\affiliation{Dipartimento di Fisica, Universit\`{a} di Roma ``Sapienza'', P.le A. Moro 2, 00185 Roma, EU}

\author{Antonino Marcian\`o}
\email{amarcian@haverford.edu}
\affiliation{Centre de Physique Th\'eorique, Case 907 Luminy, 13288 Marseille, EU}
\affiliation{Dipartimento di Fisica, Universit\`{a} di Roma ``Sapienza'', P.le A. Moro 2, 00185 Roma, EU}
\affiliation{Department of Physics and Astronomy, Haverford College, Haverford, PA 19041, USA}

\begin{abstract}
We derive the cosmological Big Bounce scenario from the dipole approximation of Loop Quantum Gravity. We show that a non-singular evolution takes place for any matter field
 and that, by considering a massless scalar field as a relational clock for the dynamics, the semi-classical proprieties of an initial state are preserved on the other side of the bounce. This model thus enhances the relation between Loop Quantum Cosmology and the full theory.  
\end{abstract}

\maketitle

\section{Introduction}
Loop Quantum Cosmology (LQC) is the most successful application of Loop Quantum Gravity (LQG). It stands as an implementation of quantization techniques adopted within LQG to specific cosmological models. In these cases classical singularities have been successfully removed \cite{Boj01,Ash06,Sin} and implications to dynamics have been studied at different levels getting 
 into Planck scale physics \cite{Ashrev,Bojrev}. Results regarding geometric inflation \cite{Bojin}, suppression of chaoticity \cite{Bojca}, addition of some inhomogeneous effects \cite{inh1,inh2}, path integral formulation within LQC \cite{sfc1}, spin foam cosmology \cite{sfc2} and relation to non-commutative geometry \cite{Bat} have been also obtained. Although these great successes, many points need still to be addressed in a more detailed analysis. For instance, the main dynamical effects have been introduced into effective classical equations or recurring to the minimal area gap argument. Despite these efforts, it is not clear how the link between LQC and the full theory can be precisely obtained \cite{CiaMon,Eng,BojKas}.

To solve these shortcomings, it has been proposed a dipole $SU(2)$ lattice theory representing a finite dimensional truncation of LQG \cite{RovVid}. As we have shown in \cite{BMR}, this {\it dipole cosmology} describes the space-time of the Bianchi IX Universe perturbed by some inhomogeneous degrees of freedom. The classical dynamics is described via the effective-like equations of LQC and a discrete quantum evolution arises without heuristic arguments. This model is at the ground of a spinfoam cosmology \cite{sfc2}.

We here complete the previous analysis by investigating the dynamics of the isotropic sector of the model, which describes the triangulation of the closed Friedman-Robertson-Walker (FRW) Universe. Our analysis is performed at classical and quantum levels showing how the cosmological singularity is tamed and a Big Bounce evolution naturally takes place. This result corroborates the Big Bang resolution previously obtained in LQC and represents a way for merging LQC in LQG. Furthermore, it provides the first derivation of a non-singular cosmological dynamics directly from LQG.

We focus on Euclidean gravity and we set the Immirzi parameter $\beta=1$. We adopt units such that $8 \pi G/3=1=c=\hbar$. 

\section{Dipole cosmology}
We present here a brief account for the model \cite{BMR} (for a concise presentation see \cite{Mar}). The basic idea is to triangulate the Cauchy surfaces slicing a generic compact homogeneous space-time. These surfaces are topologically three-spheres $S^3$. The Bianchi IX Universe is the most general model possessing such a topology, while the closed FRW Universe is the corresponding isotropic case. A topological three-sphere can be constructed by gluing together the boundaries of two three-balls and the ``equator'' of the three-sphere corresponds to the boundaries of the two balls. It follows that $S^3$ can be triangulated in the easiest way by gluing together two tetrahedra along all their faces. For a generic cellular complex triangulation $\Delta_n$ on $S^3$ involving $n$ tetrahedra $t$, one can associate a group element $U_f\in SU(2)$ and a $su(2)$ algebra element $E_f=E_f^I\tau_I$ to each oriented triangle $f$ ($\tau_I$ are $SU(2)$ generators). The phase space is then the cotangent bundle of $SU(2)^{2n}$ with its natural symplectic structure. The dynamics of the model is encoded in the two sets of constraints
\ba\label{clgau} 
G_t &=&  \sum_{f\in t} E_f = 0\, ,  \\\label{clham}
\mathcal H_t &=&  V^{-1}_t \sum_{ff'\in t} {\rm Tr}[U_{\!f\!f'}E_{f'}E_{f},] = 0\,, 
\ea
where $V_t^2=  {\rm Tr}[E_fE_{f'}E_{f'\!'}]$. The scalar constraint (\ref{clham}) provides the time evolution of the system, while (\ref{clgau}) denotes the Gau{\ss} constraint.

We restrict our attention to the dipole cosmology model, {\it i.e.} the triangulation formed by two tetrahedra defined by the dual graph $\Delta_2^*=\!\!\!\!\begin{picture}(20,5)
\put(20,4) {\circle{15}}
\put(12,4) {\circle*{2}} 
\put(28,4) {\circle*{2}}  
\qbezier(12,4)(19,13)(28,4)
\qbezier(12,4)(19,-5)(28,4)
\end{picture} \,\,\,\,\,\,$. This represents a finite dimensional truncation of LQG and describes the Bianchi IX Universe plus six inhomogeneous degrees of freedom \cite{BMR}. The $SU(2)$ symmetry structure enters twice in our description: in adding inhomogeneities and in discretizing the Ashtekar-Barbero variables. 

In terms of the homogeneous reference triad fields, the spatial slices of Bianchi IX are characterized by the Maurer-Cartan (flat) connections $\omega=g^{-1}dg=\omega^{I} \tau_I$, $\forall g\in SU(2)$, and by the dual vector fields $e^I$ taking values in $su(2)$. The Maurer-Cartan connections fulfill the $SU(2)$ structure equation $d\omega^{I}=\f12\,{\epsilon^I}_{JK}\omega^J \wedge \omega^K$ and $e^I$ the corresponding Lie brackets. By means of the first order formalism, the Ashtekar-Barbero variables for the Bianchi models are expressed as $A^I=c_{I}\, \omega^{I}$ and as $E_I=p^{I}\det(\omega^I)\, e_{I}$, where $c_I=c_I(t)$ and $p^I=p^I(t)$ are conjugate variables describing the dynamics \cite{Bojrev}.

Once discretized, the flux of the electric field rewrites exactly as $E_f=p^I\,\omega^I_f\tau_I$, while the holonomy around the dual links of $\Delta_2^*$ can be approximated as $U_f=\exp\,c^I\,\omega^I_f\tau_I$. The $\omega^I_f$ are the building blocks of the model. They are defined as the circuitations along the dual links of $\Delta_2^*$ and represent the flux of the Plebanski two-form $\Sigma^I=\f12\,{\epsilon^I}_{JK}\omega^J \wedge \omega^K$ across triangles of $\Delta_2^*$. By means of Maurer-Cartan structure equation, they read 
\be
\omega_{f}^I \equiv  \int_f \Sigma^I = \int_f  d \omega^I = \oint_{\partial f}  \omega^I\,.
\ee
Thus the Gau{\ss} constraint (\ref{clgau}) is satisfied.
 
If we restrict our attention to the homogeneous isotropic sector of the dipole model, the phase space becomes two-dimensional and coordinatized by $(c,p)$. The first order formalism is related to the metric one via the relations $|p| =  a^2$ and $c = (1+\dot{a})/2$, where $a=a(t)$ denotes the scale factor of the closed FRW Universe. The classical dynamics of this model is described by the Hamiltonian constraint \cite{BMR} 
\begin{equation}\label{coniso} 
\mathcal H_g=\frac{17}{6}\sqrt p\,\left(\cos(c-\alpha)-1\right)=0\,,
\end{equation}
where $\alpha$ takes into account the contribution of the $S^3$ curvature to the holonomy in the expression $U_f=\exp(c+\alpha)\omega^I_f\tau_I$. The value of $\alpha$, such that $\cos\alpha=(9-\sqrt{17})/8$, is fixed by requiring matching with ordinary classical dynamics for low curvature ($|c|\ll1$). The scalar constraint (\ref{coniso}) resembles the effective constraint of LQC \cite{Tav}, but is here obtained without polymerizing the classical theory.
 
The quantization of the triangulated isotropic model is straightforward \cite{BMR}. The kinematic Hilbert space of the theory is $L^2(S^1,dc/4\pi)$, since the variable $c$ multiplies a generator of a $U(1)$ subgroup of $SU(2)$. Eigenstates of $p$ are labeled by integer $\mu$ and read $\langle c|\mu\rangle=e^{i\mu c/2}$. Wave functions $\psi(c)$ are decomposed in a Fourier series of eigenstates of $p$. The fundamental operators are $p$ and $\exp(i c/2)$ and their action reads $p\, |\mu\rangle = \mu/2\ |\mu\rangle$ and $\exp(i c/2)\,|\mu\rangle=|\mu+1\rangle$. The quantum constraint operator then rewrites as a difference equation as in standard LQC \cite{Ashrev,Bojrev}. However, the discrete dynamics is here recovered without recurring to the minimal area gap argument.

\section{Classical dynamics}
We now analyze the classical dynamics of the system described by the constraint (\ref{coniso}) in the presence of a generic matter field. Once  (\ref{coniso}) it is taken into account, the total Hamiltonian for the system reads
\be
\mathcal H=\mathcal{H}_{g}+\mathcal{H}_{m}=\frac{17}{6} \sqrt{p}\, ( \cos(c-\alpha) -1 ) +|p|^{3/2}\rho\,,
\ee
where $\rho=\rho(t)$ is the matter energy density. The equations of motion are given by 
\ba \label{cpunto}
\dot{c}&=& \frac{17}{12} \frac{N}{\sqrt{p}} \left( \cos(c-\alpha)-1 \right) + \frac{3}{2}N\sqrt{p}\,\rho  + N p^{3/2} \rho'\,,\,\,\,\,\\\label{ppunto}
\dot{p}&=& \frac{17}{6} N  \sqrt{p}\,  \sin (c-\alpha)\,,
\ea
in which $N$ denotes the lapse function and $\rho'=d\rho/ d p$. On the other hand, enforcing the scalar constraint $\mathcal{H}=0$, the relation
\be\label{relcon}
\cos(c-\alpha)=1-\f{6}{17}\, p\,\rho
\ee
holds. The equation of motion for the Hubble rate $(\dot{a}/a)$ can be then obtained by substituting (\ref{relcon}) into (\ref{ppunto}) and reads
\be\label{modfri} 
\left(\f{\dot a}a\right)^2=\left( \frac{\dot{p}}{2 p} \right)^2=\f{17}6N^2\rho\left(1-\f\rho{\rho_c}\right)\,,
\ee
where we have defined $\rho_c=17/(3p)$. This is the Friedman equation for the isotropic dipole cosmology and it clearly exhibits a non-singular dynamics. The difference with respect to the standard dynamics relies in the $\rho^2$-term which, as soon as the Universe expands enough ($|p|\gg1$), is negligible. In fact, as for ordinary matter fields\footnote{We recall that the energy density $\rho$ behaves as \cite{PriCos}: $\rho\propto|p|^{-3/2}$ for a cosmological constant term; as $\rho\propto|p|^{-2}$ for ultra-relativistic particles; as $\rho\propto|p|^{-5/2}$ for a perfect gas and as $\rho\propto|p|^{-3}$ for a free scalar field.} $\rho\sim|p|^{-\gamma}$ with $\gamma>1$, the term $p\rho$ vanishes as $|p|\gg1$. The standard behavior of the Friedman Universe is then recovered for a large scale factor. Notice that, as expected, the low curvature limit ($|c|\ll1$) corresponds to the large volume limit ($|p|\gg1$). On the other hand, the $\rho^2$-factor is relevant in high energy regime ($|p|\ll1$), {\it i.e.} in the Planck era. As the energy density $\rho$ reaches the critical value $\rho_c$, the Hubble rate vanishes and the Universe experiences a bounce (or more generally a turn-around) in the scale factor. This result agrees with LQC\footnote{Notice that our $\rho_c$ depends on the scale factor as in the so-called old quantization scheme \cite{Ash06a}. The relation between the old and the improved dynamics in LQC is described in \cite{CorSin08}. The improved quantization of the dipole cosmology will be reported in \cite{BatMar}.} \cite{Ash06,Sin,Ashrev,Bojrev}.

Let us now analyze the Raychaudhuri cosmological equation, {\it i.e.} that describes the acceleration of the Universe. Considering the above equations of motion (\ref{cpunto}) and (\ref{ppunto}), and taking into account the relation (\ref{relcon}), such an equation takes the form
\be\label{rayeq}
\frac{\ddot{a}}{a}=\frac{\ddot{p}}{2p}-\left(\frac{\dot{p}}{2p}\right)^2=\f{N^2}{12}p\left(\f{\rho'}{17}-3\rho^2-6p\,\rho\rho'\right)\,.
\ee
The last two terms are negligible as large regions of the Universe are considered. In fact, as explained above, these two terms behave as $\rho^2\sim p\,\rho\rho'\sim|p|^{-2\gamma}$ and are then negligible with respect to $\rho'$ as $|p|\gg1$. The correct Friedman limit is thus recovered. By considering the cosmological constant term $\rho=\Lambda/ p^{3/2}$, it follows that for $|p|\gg1$ we have $\ddot{a}\,a^2= -\Lambda N^2/8<0$. This result is consistent with the $\Lambda$-Cold-Dark-Matter cosmological model \cite{PriCos}. 

To put forward the comparison with respect to LQC, we consider the dynamics in the presence of a massless scalar field $\phi$. Its energy density is $\rho_\phi=p_\phi^2/|p|^3$, where $p_{\phi}$ is the conjugate momentum to $\phi$. Its Hamiltonian density, in a homogeneous isotropic Universe, is characterized by the only kinetic term $\mathcal{H}_{\phi}=p^2_{\phi}/|p|^{3/2}$ and the absence of a potential implies $p_{\phi}$ to be constant. The total Hamiltonian constraint for this model is given by 
\be\label{htot}
\mathcal H=\mathcal H_g+\mathcal H_\phi=\frac{17}{6} \sqrt{p}\, ( \cos(c-\alpha) -1 ) +\f{p^2_{\phi}}{|p|^{3/2}}=0
\ee
and the corresponding Friedman equation then reads
\be\label{deffri}
\left(\frac{\dot{p}}{2p}\right)^2 = \frac{17}{6} N^2 \f{p_{\phi}^2}{|p|^{3}}\left(1-\frac{3}{17}\f{p_\phi^2}{p^{2}}  \right)\,.
\ee
The phase space of this model is four-dimensional, with coordinates $(c,p,\phi,p_\phi)$ and, since $p_{\phi}$ is a constant of motion, each classical trajectory can be specified in the $(p,\phi)$-plane, namely in the $(a,\phi)$-plane. In other words, the scalar field $\phi$ can be regarded as a (relational) clock for the Universe dynamics\footnote{We underline that the use of a scalar field as a relational clock is not a necessary tool of this analysis. A consistent and phisically motivated relational description of the cosmological system can be developed. Anyway, in order to make contact with the existent literature, we adopt such a framework, postponing relational issues to \cite{fut}.}. This condition can be imposed by fixing the lapse function to satisfy the time gauge $\dot\phi=1$. More explicitly, we have
\be\label{lapsefix}
\dot{\phi}= N \frac{\partial \mathcal{H}}{\partial p_{\phi}}= \frac{2 N p_{\phi}}{|p|^{3/2}}=1 \quad \Rightarrow \quad N=\f{|p|^{3/2}}{2p_\phi}\,.
\ee
In this case, the deformed Friedmann equation (\ref{deffri}) rewrites as 
\be\label{friphi}
\left( \frac{1}{2p}\,\frac{dp}{d \phi} \right)^2 = \frac{17}{24} \left(1-\frac{3}{17}\frac{p_\phi^2}{p^2}\right)\,,
\ee
whose solution is given by $p(\phi)\sim e^{-2\sqrt{17/24}\,\phi}(\tilde{p}_\phi^2+e^{4\sqrt{17/24}\,\phi})$, where $\tilde p_\phi^2=3p_\phi^2/17$. The equation (\ref{friphi}) implies a Big Bounce dynamics and the scale factor shows a minimum non vanishing value in $a_{\text{min}}=\sqrt{\tilde p_\phi}$. The standard solutions $a(\phi)\sim e^{\pm\sqrt{17/24}\,\phi}$ are recovered at late times $|\phi|\rightarrow\infty$, namely as large scale regions ($p\gg\tilde p_\phi$) are taken into accounts.

Finally, within the $\phi$-evolutionary scheme, the Raychaudhuri equation (\ref{rayeq}) for the system (\ref{htot}) takes the form
\be\label{rayeqphi} 
\frac{\ddot{a}}{a}= \frac1{16}\left(-\f1{17}+11\,\f{p_\phi^2}{p^2}\right)\,.
\ee
At the bouncing point $p=\tilde p_\phi$, the acceleration of the Universe is greater then zero. Then, as soon as the Universe expands enough ($p\gg\tilde p_\phi$), the second term in the right hand side of (\ref{rayeqphi}) becomes negligible and the dynamics is governed by a negative acceleration.

\section{Quantum dynamics}
In this part we discuss the quantum dynamics of the system (\ref{htot}). The quantization of the model has been discussed in \cite{BMR} and above recalled. Once eigenfunctions from the quantum Hamiltonian constraint $\mathcal{H}\Psi=0$ have been recovered, the expectation value of any operators $\mathcal{O}$ can be evaluated on a wave packet which is constructed over the space of solutions of the quantum constraint. That means to compute $\langle\dot{\mathcal{O}}\rangle= -i \langle \left[ \mathcal{O}, \mathcal{H}_{e} \right]\rangle$, where $\mathcal H_e$ denotes an effective Hamiltonian (see below). In particular, we are interested in evaluating the mean values $\langle\dot{c}\rangle$ and $\langle\dot{p}\rangle$, as well as the evolution of the scale factor relative spreading $(\Delta p)^2/\langle p\rangle^2$, where $(\Delta p)^2=\langle p^2\rangle-\langle p\rangle^2$. It is relevant to analyze the relative spreading in order to investigate the semi-classical proprieties of the quantum Universe. To be more precise, an observable $\mathcal O$ is called semi-classical if two conditions are satisfied: (i) its expectation value is close to the classical one along the whole dynamical evolution; (ii) its relative fluctuations are small, {\it i.e.} $(\Delta\mathcal O)^2/\langle\mathcal O\rangle^2\ll1$.

Before quantizing, we rewrite the constraint (\ref{htot}) at the classical level in such a way to assume a Schr\"odinger-like form. Indeed, by fixing the lapse function as in (\ref{lapsefix}), the momentum $p_{\phi}$ has been considered as the generator of $\phi$-time-translations and thus we deal with an effective Hamiltonian $p_{\phi}\equiv \mathcal{H}_{e}$:
\be
\mathcal H_e=\sqrt{\frac{17}{6}}\,p\,\sqrt{1- \cos (c- \alpha)}\,.
\ee
This is a true Hamiltonian (not yet a constraint) in relation to which all observables evolve. Within the previously recalled quantization scheme, we obtain the equations of motion for the mean values 
\begin{eqnarray}\label{rayqua}
\f d{d\phi}\langle c\rangle &=& \left\langle\f{p_\phi}p\right\rangle\,,\\\label{friqua}
\f d{d\phi}\langle p\rangle &=& \sqrt{\frac{17}{6}}\left\langle p\,\sqrt{1- \frac{3}{17} \left( \frac{p_{\phi}}{p} \right)^2}\right\rangle\,,
\end{eqnarray}
where we have used once again the constraint $\mathcal{H}=0$. The quantum trajectories (\ref{rayqua}) and (\ref{friqua}) are in exact agreement with the classical ones. In particular, (\ref{friqua}) corresponds to the Friedman equation (\ref{friphi}), while (\ref{rayqua}) is related to the Raychaudhuri one (\ref{rayeqphi}). As the Universe reaches large scale factor regions ($|p|\gg1$) the ordinary solutions are recovered.

Let us now analyze the evolution of the scale factor relative fluctuations $(\Delta p)^2/\langle p\rangle^2$. The $\phi$-time evolution of the spreading $(\Delta p)^2$ is not constant and neither bounded already in the ordinary case as $(\Delta p)^2\sim e^{\phi}$. On the other hand, the scale factor relative fluctuations are governed by the equation
\begin{multline}\label{relflu}
\frac{d}{d\phi}\left(\f{(\Delta p)^2}{\langle p\rangle^2}\right)= \sqrt{\frac{17}{3}} \frac{1}{\langle p\rangle^2} \left( \left\langle p^2  \sqrt{1- \frac{3}{17} \left( \frac{p_{\phi}}{p} \right)^2}  \right\rangle\right.+\\ - \left.\frac{ \langle p^2\rangle}{\langle p\rangle} \left\langle p  \sqrt{1- \frac{3}{17} \left( \frac{p_{\phi}}{p} \right)^2}  \right\rangle  \right)\,,
\end{multline}
and, in the Friedman framework, such a quantity is conserved during the whole evolution. The semi-classicality of an initial state is then preserved during the dynamics. On the other hand, in the triangulated framework, this propriety is valid for large scale factor $|p|\gg1$ only (namely, at late times $|\phi|\rightarrow\infty$). As the scale factor reaches its minimum value, the quantity (\ref{relflu}) vanishes. 

We can extract relevant informations about the semi-classical properties of the triangulated Universe by analyzing the difference in the asymptotic values 
\be\label{d}
\mathcal{D}=\left|\left(\f{(\Delta p)^2}{\langle p\rangle^2}\right)_{\phi\rightarrow-\infty}-\left(\f{(\Delta p)^2}{\langle p\rangle^2}\right)_{\phi\rightarrow\infty}\right|\,.
\ee
For this purpose we analyze the Wheeler-De Witt (WDW) equation for the model (\ref{htot}). By considering $\phi$ as the time coordinate, this equation reads
\begin{equation}\label{wdweq}
\left(\partial_{\phi}^2+\mathcal H_e^2\right)\Psi=0, \quad \mathcal H_e^2= - \frac{17}{6} \left( 1- \cos(c-\alpha) \right) \partial_{c}^2
\end{equation}
where $\Psi=\Psi(c,\phi)$. As usual the WDW equation can be thought as a Klein-Gordon-like equation where $\phi$ plays the role of time and $\mathcal H_e^2$ of the spatial Laplacian. In order to have an explicit Hilbert space, we perform the natural frequencies decomposition of the solution of (\ref{wdweq}) and focus on the positive frequency sector. The wave function $\Psi_\omega(c,\phi)=e^{i\omega\phi}\psi_\omega(c)$ is of positive frequency with respect to $\phi$, where $\omega^2$ denotes the spectrum of $\mathcal H_e^2$. It satisfies the positive frequency square root of the quantum constraint (\ref{wdweq}). We are then dealing with the Schr\"odinger-like equation $-i\p_\phi\Psi=\mathcal H_e\Psi$, corresponding to the Hamiltonian system described by $p_\phi=\mathcal H_e$. A generic wave packet solution reads
\ba\label{wavepa}
\Psi(c, \phi)&=& \int d\mu_\omega\, e^{i \omega \phi} \psi_{\omega}(c)\,,\\\nonumber
\psi_\omega(c)&=&\psi_0\exp\left(i\, \omega \sqrt{\frac{6}{17} } \int_{c_0}^c \frac{d c'}{\sqrt{1- \cos(c'-\alpha)}}\right)\,,
\ea
where $d\mu_\omega=A(\omega)\,d\omega$ denotes the measure, $A(\omega)$ being the weighting function. It is worth stressing that the wave function satisfies the property $\Psi_\omega(c,\phi)=\Psi^*_{-\omega}(c,\phi)$, in which $\phantom{}^*$ denotes complex conjugation. The mean value of an operator $\mathcal{O}$ is given by $\langle \mathcal{O}\rangle(\phi)=\int d\mu_\omega\,d\mu_\omega' \, e^{i (\omega-\omega') \phi} \langle\psi^*_{\omega'} \mathcal{O}\, \psi_{\omega} \rangle$, and since for any self-adjoint operator $\mathcal{O}$ the property $\langle\psi^*_{\omega'} \mathcal{O}\, \psi_{\omega} \rangle=\langle\psi^*_{\omega} \mathcal{O}\, \psi_{\omega'} \rangle$ holds, we get the relation
\be
\langle \mathcal{O}\rangle(-\phi)= \int d\mu_\omega\,d\mu_\omega' \, e^{-i (\omega-\omega') \phi} \langle\psi^*_{\omega'}\mathcal{O}\psi_{\omega} \rangle=\langle \mathcal{O}\rangle(\phi) \,.
\ee
This way, the mean value of any self-adjoint operator $\mathcal O$ with respect to the states (\ref{wavepa}) is invariant under time inversion $\phi\rightarrow-\phi$. A natural choice for $d\mu_\omega$ is to consider a Gau{\ss}ian weighting function picked around $\omega_0$ with standard deviation $\sigma$, {\it i.e. }  $A(\omega)=e^{-(\omega-\omega_0)^2/{2 \sigma^2}}$.

The quantity (\ref{d}) is thus vanishing as both fluctuations $(\Delta p)^2(\phi)$ and mean value $\langle p\rangle(\phi)$ are symmetric in time. Therefore, although the relative fluctuations $(\Delta p)^2/\langle p\rangle^2$ are in general not constant during the dynamics (see equation (\ref{relflu})), the semi-classicality on an initial state is preserved on the other side of the Big Bounce. More precisely, if we start the evolution by considering a Gau{\ss}ian semi-classical state such that $(\Delta p)^2/\langle p\rangle^2\ll1$ at large volumes, this propriety will be satisfied on the other side of the bounce when the Universe approaches large scales ($p\gg1$). This feature is in agreement with the so-called cosmic recall in LQC \cite{CosRec1,CosRec2}. 

\section{Concluding remarks}
In this paper we have derived a bouncing cosmology from the dipole approximation of LQG. A coarse triangulation of the physical space has been fixed and we have considered a discretization and quantization of gravity on this triangulation. This model shows a Big Bounce for the closed FRW Universe without recurring to polymerization or minimal area gap arguments usually invoked in LQC. The link between LQC and LQG is then improved.

Progresses in this line of research should be the generalization to the case of Immirzi parameter $\beta\neq 1$, to Lorentzian signature and the corresponding analysis of the Belinski-Khalatnikov-Lifshitz scenario in this latter case.

\section*{Acknowledgments}

We thank F. Vidotto for her valuable comments.

\end{document}